\documentclass[twocolumn,showpacs,amsmath]{revtex4}%
\usepackage{amsfonts}
\usepackage{amsmath}
\usepackage{amssymb}
\usepackage{graphicx}%
\providecommand{\U}[1]{\protect\rule{.1in}{.1in}}

\begin{document}
\title[ ]{Narrowband Biphoton Generation due to Long-Lived Coherent Population Oscillations}
\author{A.V. Sharypov}
\affiliation{Department of Chemistry, Bar-Ilan University, Ramat Gan 52900, Israel}
\author{A.D. Wilson-Gordon }
\affiliation{Department of Chemistry, Bar-Ilan University, Ramat Gan 52900, Israel}
\author{}
\affiliation{}

\begin{abstract}
We study the generation of paired photons due to the effect of four-wave
mixing in an ensemble of pumped two-level systems that decay via an
intermediate metastable state. The slow population relaxation of the
metastable state creates long-lived coherent population oscillations, leading
to narrowband nonlinear response of the medium which determines the spectral
width of the biphotons. In addition, the biphotons are antibunched, with
antibunching period determined by the dephasing time. During this period,
damped oscillations of the biphoton wavefunction occur if the pump detuning is non-zero.

\end{abstract}
\date{\today}
\maketitle







\bigskip Traditionally, paired photons are produced from spontaneous
parametric down conversion in nonlinear crystals. The bandwidth of such
biphotons is very broad and typically in the terahertz range
\cite{YarivBook2007}, which makes them useless for some applications in
quantum information science which require strong interaction between photons
and atomic systems. This problem can be overcome by generating biphotons in
cold atomic systems which have a narrowband nonlinear response. For example,
biphotons can be produced when a double-$\Lambda$ system
\cite{NBHarris,NB,NBTheoryEIT} or two-level system (TLS) \cite{NBOld,NBTLS} is
pumped by two counter-propagating laser fields. Then phase-matched and
energy-time entangled photon pairs are produced due to the effect of four-wave
mixing (FWM). Biphotons from a such source have a bandwidth in the megahertz
range and coherence time of hundreds of nanoseconds.

In this paper, we demonstrate that narrowband biphotons can also be produced
due to the effect of long-lived coherent population oscillations (CPOs) in a
TLS with an intermediate metastable state. In such a system, the width of the
nonlinear response is determined by the lifetime of the metastable state,
which can vary significantly depending on the nature of the quantum system.
For example, in semiconductor quantum wells and dots \cite{CPOQdots} the CPO
lifetime is in the microsecond range, whereas in a ruby crystal \cite{CPORuby}
or organic film \cite{CPOorg} it can be more than a millisecond, leading to a
broad range of potential applications of photon pairs based on the CPO effect.

\begin{figure}
[ptb]
\begin{center}
\includegraphics[
trim=0.000000in 0.000000in 0.457373in 0.308923in,
height=2.0678in,
width=3.3122in
]%
{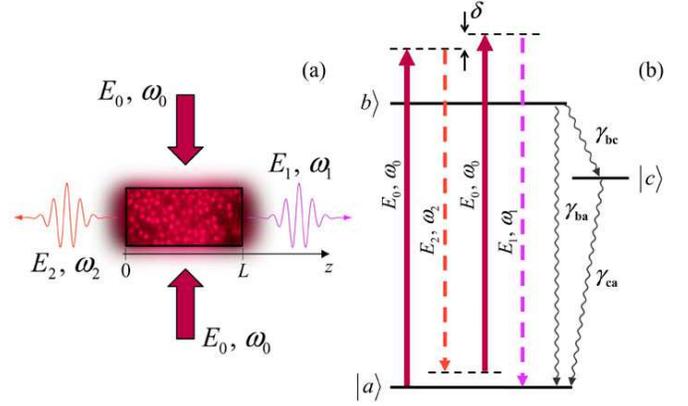}%
\caption{(color online). In the presence of the counter-propagating pump
fields, phase-matched counter-propagating biphotons are generated inside the
medium due to FWM.}%
\label{Fig1}%
\end{center}
\end{figure}

We consider the interaction of an ensemble of TLSs that decay via a single
intermediate metastable state with two counterpropagating pump fields with
amplitude $E_{0}$\ (see Fig.\ref{Fig1}). The medium is assumed to be optically
thin in the direction of pump propagation and the effect of pump depletion is
not taken into account. Due to pumping of the TLS by two counter-propagating
laser fields, photon pairs are produced \cite{NBOld,NBTLS} where photons from
the same pair also counter-propagate so that the phase-matching condition of
FWM is satisfied \cite{BoydBook2003} [see Fig. \ref{Fig1}(a)] (actually, in
such a configuration biphotons are emitted into the whole $4\pi$ space). In
order to allow for the spontaneous initiation process, the generated weak
fields are described by quantum-mechanical operators%
\begin{equation}
E_{1,2}^{\left(  +\right)  }=E_{1,2}^{\left(  0\right)  }\widehat{a}%
_{1,2}\left(  z,t\right)  ~\text{and }E_{1,2}^{\left(  -\right)  }%
=E_{1,2}^{\left(  0\right)  }\widehat{a}_{1,2}^{\dagger}\left(  z,t\right)  ,
\label{01 QuantFieldsNot}%
\end{equation}
where the subscript $1$ denotes the field at frequency $\omega_{1}=\omega
_{0}+\delta$ propagating along the positive z-axis, and the subscript $2$
denotes the field at frequency $\omega_{2}=\omega_{0}-\delta$ moving along the
negative z-axis, $E_{1,2}^{\left(  0\right)  }=\left(  \hbar\omega
_{1,2}/2\varepsilon_{0}V\right)  ^{1/2}$ is the vacuum field, $V$ is the
quantization volume, and $\widehat{a}$ and $\widehat{a}^{\dagger}$ are the
photon annihilation and creation operators.\bigskip%

To describe the evolution of the atomic ensemble, we begin with the Heisenberg
operator equations of the motion in the dipole approximation:%
\begin{align}
\hbar\left(  d/dt+\Gamma_{ba}+i\omega\right)  \widetilde{\sigma}_{ba} &
=id_{ba}\widetilde{E}^{\left(  +\right)  }\left(  \widetilde{\sigma}%
_{bb}-\widetilde{\sigma}_{aa}\right)  ,\nonumber\\
\hbar\left(  d/dt+\Gamma_{ba}-i\omega\right)  \widetilde{\sigma}_{ab} &
=-id_{ab}\widetilde{E}^{\left(  -\right)  }\left(  \widetilde{\sigma}%
_{bb}-\widetilde{\sigma}_{aa}\right)  ,\label{02 DensMatrix}\\
\hbar\left(  d/dt+\gamma_{b}\right)  \widetilde{\sigma}_{bb} &  =id_{ab}%
\widetilde{E}^{\left(  -\right)  }\widetilde{\sigma}_{ba}-id_{ba}\widetilde
{E}^{\left(  +\right)  }\widetilde{\sigma}_{ab},\nonumber\\
\left(  d/dt+\gamma_{ca}\right)  \widetilde{\sigma}_{cc} &  =\gamma
_{bc}\widetilde{\sigma}_{bb},\nonumber
\end{align}
where $\widetilde{\sigma}_{ij}=\left\vert i\right\rangle \left\langle
j\right\vert $ is the atomic operator, $\widetilde{E}^{\left(  \pm\right)  }$
is the total field operator, and $d_{ij}$ is the transition dipole matrix
element, $\Gamma_{ba}$ is the transverse relaxation rate, $\gamma_{ij}$ is the
longitudinal decay rate from the state $\left\vert i\right\rangle $ to the
state $\left\vert j\right\rangle $, $\gamma_{b}=\gamma_{bc}+\gamma_{ba}$ is
the total decay rate from the excited level, and $\Omega=\omega_{0}%
-\omega_{ba}$ is the pump detuning from the resonance. We also assume that the
system is closed so that $\widetilde{\sigma}_{aa}+\widetilde{\sigma}%
_{bb}+\widetilde{\sigma}_{cc}=1.$

We apply the slowly-varying envelope approximation and write the total field
operator as%
\begin{equation}
\widetilde{E}^{\left(  \pm\right)  }=\left(  E_{0}+E_{1}^{\left(  \pm\right)
}e^{\mp i\delta t}+E_{2}^{\left(  \pm\right)  }e^{\pm i\delta t}\right)
e^{\mp i\omega_{0}t}. \label{03 RWA Fields}%
\end{equation}

\noindent To eliminate the fast oscillating term in Eqs. (\ref{02 DensMatrix}%
), we introduce the transformations $\widetilde{\sigma}_{ba,ab}\left(
t\right)  =\sigma_{ba,ab}\left(  t\right)  e^{\mp i\omega_{0}t}$ and
$\widetilde{\sigma}_{jj}\left(  t\right)  =\sigma_{jj}\left(  t\right)
|_{j=a,b,c}$. In order to find the medium response to the weak generated
fields, we apply the Floquet theory \cite{TannorBook} and write%
\begin{equation}
\sigma=\sigma^{\left(  0\right)  }+\sigma^{\left(  +\delta\right)
}e^{-i\delta t}+\sigma^{\left(  -\delta\right)  }e^{i\delta t}.
\label{04 Floquet}%
\end{equation}

\noindent The zeroth-order solution of Eqs. (\ref{02 DensMatrix}) -
(\ref{04 Floquet}) gives the response of the medium to the pump field and the
population distribution between the quantum states, whereas the first-order
solution determines the medium response to the weak generated fields since
$P_{1,2}=Nd_{ba}\sigma_{ba}^{\left(  \pm\delta\right)  }$ \cite{CPO}.

The pump-probe interaction with a TLS is characterized by population beating
or coherent population oscillations (CPOs) at $\delta$, the frequency
difference between pump and probe fields \cite{CPO}. In an ordinary TLS, the
CPOs decay at the same rate as the excited state. However, the situation can
be quite different if an intermediate metastable state is included [see Fig.
\ref{Fig1}(b)]. In the case where the rate of transfer of CPO from the excited
state to the metastable state is much faster than the rate of transfer of CPO
from the excited to the ground state due to population relaxation or
pump-induced transitions, that is,%
\begin{equation}
\gamma_{bc}\gg V_{0},\gamma_{ba}, \label{05 CPO condition}%
\end{equation}
where $V_{0}=d_{ba}E_{0}/\left(  2\hbar\right)  $ is the pump Rabi frequency
which is assumed to be real, long-lived CPOs of ground and metastable states
are created \cite{AsiMemory}. This leads to a narrow dip in the probe
absorption spectrum and a narrow peak in the FWM spectrum
\cite{AsiMemory,ArleneTLS,MyCPO}.

Under these conditions and taking into account that level $\left\vert
c\right\rangle $ is the metastable level $\gamma_{ca}\ll\gamma_{ba}%
,\gamma_{bc}$ the steady-state response of the medium to the generated fields
is given by \cite{MyCPO,MyHellerCPO}%
\begin{subequations}
\begin{align}
\sigma_{ba}^{\left(  +\delta\right)  } &  =\left(  \alpha_{1}E_{1}^{\left(
+\right)  }+\beta_{1}E_{2}^{\left(  -\right)  }\right)  d_{ba}/\hbar
,\label{06a Probe Response}\\
\sigma_{ab}^{\left(  -\delta\right)  } &  =\left(  \alpha_{2}E_{2}^{\left(
-\right)  }+\beta_{2}E_{1}^{\left(  +\right)  }\right)  d_{ab}/\hbar
,\label{06b}%
\end{align}
where $\alpha_{1,2}$ are proportional to the effective linear susceptibilities
and $\beta_{1,2}$ are proportional to the effective third-order nonlinear
susceptibilities and are responsible for the generation of the paired photons.
They are given by%
\end{subequations}
\begin{subequations}
\begin{align}
\alpha_{1,2} &  \equiv\pm i\left(  1+X\right)  /\left[  \Gamma_{1,2}\left(
1+\kappa\right)  \right]  ,\label{07a linear response}\\
\beta_{1,2} &  \equiv\pm iX/\left[  \Gamma_{1,2}\left(  1+\kappa\right)
\right]  ,
\end{align}
where%
\end{subequations}
\begin{subequations}
\begin{equation}
X\equiv-\kappa\gamma_{ca}/\left(  W-i\delta\right)  \label{08a coherence term}%
\end{equation}
is the coherent field interaction term and%
\begin{equation}
W=\left(  1+\kappa\right)  \gamma_{ca}\label{08b CPO width}%
\end{equation}
determines the characteristic width of the window in which coherent
interaction between the fields occurs, $\kappa\equiv2V_{0}^{2}/\left[
\gamma_{ca}\Gamma_{ba}\left(  1+\Omega^{2}/\Gamma_{ba}^{2}\right)  \right]  $
is the saturation parameter, and $\Gamma_{1,2}\equiv\Gamma_{ba}+i\left(
\mp\Omega-\delta\right)  .$

The evolution of the annihilation and creation operators $\widehat{a}$ and
$\widehat{a}^{\dagger}$ is described by the coupled propagation equations
\cite{NBTheoryEIT}:%
\end{subequations}
\begin{subequations}
\begin{align}
\left(  \frac{\partial}{\partial t}+c\frac{\partial}{\partial z}\right)
\widehat{a}_{1}\left(  z,t\right)   &  =ig_{1}N\sigma_{ba}^{\left(
+\delta\right)  },\label{09a prop Eq}\\
\left(  \frac{\partial}{\partial t}-c\frac{\partial}{\partial z}\right)
\widehat{a}_{2}^{\dagger}\left(  z,t\right)   &  =-ig_{2}N\sigma_{ab}^{\left(
-\delta\right)  }, \label{09b}%
\end{align}
where $g_{1,2}=d_{ba}E_{1,2}^{\left(  0\right)  }/\hbar$ are the coupling constants.

In order to solve these coupled equations we make a Fourier transformation,
neglect the term $i\omega/c$ as it does not affect the final result
\cite{NBTheoryEIT}, and substitute Eqs. (\ref{06a Probe Response}) and
(\ref{06b}) into Eqs. (\ref{09a prop Eq}) and (\ref{09b}) to obtain%
\end{subequations}
\begin{subequations}
\begin{align}
\partial\widehat{a}_{1}/\partial\left(  \zeta z\right)   &  =i\alpha
_{1}\widehat{a}_{1}+i\beta_{1}\widehat{a}_{2}^{\dagger}%
,\label{10a Prop Eq Fourier}\\
\partial\widehat{a}_{2}^{\dagger}/\partial\left(  \zeta z\right)   &
=-i\alpha_{2}\widehat{a}_{2}^{\dagger}-i\beta_{2}\widehat{a}_{1},\label{10b}%
\end{align}
where $\zeta=\sigma_{cs}\Gamma_{ba}%
\mathbb{N}
/2$, $\sigma_{cs}=d_{ba}^{2}\omega_{0}/\left(  c\varepsilon_{0}\hbar
\Gamma_{ba}\right)  $ is the atomic cross section, $%
\mathbb{N}
=N/V$ is the atomic density and it is also assumed that $E_{1}^{\left(
0\right)  }\approx E_{2}^{\left(  0\right)  }$. The biphotons
counter-propagate so that photon `1' leaves the medium at the point $z=L$ and
photon `2' leaves at $z=0$. The boundary conditions derive from the vacuum
field fluctuations at $z=0$ for photon `1' and at $z=L$ for photon `2' [see
Fig. \ref{Fig1}(a)]. Thus, the solution of this system for variables
$\widehat{a}_{1}\left(  L\right)  $ and $\widehat{a}_{2}^{\dagger}\left(
0\right)  $ of the backward-wave problem can be written as a linear
combination of the initial boundary values%
\end{subequations}
\begin{subequations}
\begin{align}
\widehat{a}_{1}\left(  L\right)   &  =A_{1}\widehat{a}_{1}\left(  0\right)
+B_{1}\widehat{a}_{2}^{\dagger}\left(  L\right)  ,\label{11a Solution of Prop}%
\\
\widehat{a}_{2}^{\dagger}\left(  0\right)   &  =A_{2}\widehat{a}_{2}^{\dagger
}\left(  L\right)  +B_{2}\widehat{a}_{1}\left(  0\right)  ,
\end{align}

where%
\end{subequations}
\begin{subequations}
\begin{align}
A_{1,2}  &  =e^{\pm i\left(  \alpha_{1}-\alpha_{2}\right)  \widetilde{L}%
/2}/D,\label{12a AB const}\\
B_{1,2}  &  =i\beta_{1,2}\sin\left(  R\widetilde{L}\right)  /\left(
RD\right)  ,\label{12b}\\
D  &  =\cos\left(  R\widetilde{L}\right)  -i\frac{\alpha_{1}+\alpha_{2}}%
{2R}\sin\left(  R\widetilde{L}\right)  ,\nonumber\\
R  &  =\left[  \left(  \alpha_{1}+\alpha_{2}\right)  ^{2}/4-\beta_{1}\beta
_{2}\right]  ^{1/2}\text{ and }\widetilde{L}=\zeta L.\nonumber
\end{align}

The correlation between the photons emitted to the left and right is described
by the second-order Glauber correlation function $G_{21}^{\left(  2\right)  }$
\cite{photonsBook}%
\end{subequations}
\begin{equation}
G_{21}^{\left(  2\right)  }\left(  \tau\right)  =\left\langle \widehat{a}%
_{1}^{\dagger}\left(  L,t\right)  \widehat{a}_{2}^{\dagger}\left(
0,t+\tau\right)  \widehat{a}_{2}\left(  0,t+\tau\right)  \widehat{a}%
_{1}\left(  L,t\right)  \right\rangle . \label{15 General Cor Function}%
\end{equation}
As a field emitted by many statistically independent atoms behaves as a
Gaussian random variable, we can use the Gaussian momentum theorem
\cite{NBTheoryEIT,photonsBook} and rewrite Eq. (\ref{15 General Cor Function})
in the form%
\begin{subequations}
\begin{equation}
G_{21}^{\left(  2\right)  }=G_{1}^{\left(  1\right)  }\left(  0\right)
G_{2}^{\left(  1\right)  }\left(  0\right)  +\left\vert \Phi_{21}\left(
\tau\right)  \right\vert ^{2}, \label{16a Gauss Theorem Cor Funct}%
\end{equation}
where the terms%
\begin{equation}
G_{1,2}^{\left(  1\right)  }\left(  0\right)  =\left\langle \widehat{a}%
_{1,2}^{\dagger}\left(  0\right)  \widehat{a}_{1,2}\left(  0\right)
\right\rangle \label{16b First Order CF}%
\end{equation}
describe the appearance of uncorrelated photons which produce a flat
background, and the second term%
\begin{equation}
\Phi_{21}\left(  \tau\right)  =\left\langle \widehat{a}_{2}\left(
0,t+\tau\right)  \widehat{a}_{1}\left(  L,t\right)  \right\rangle
\label{16c Two Photon CF}%
\end{equation}
describes the appearance of entangled photon pairs and corresponds to the
biphoton wave function \cite{NBTheoryEIT,photonsBook,Rubin1994}.

As seeding fields $\widehat{E}_{1,2}$ are absent, the vacuum field
fluctuations determine the initial conditions and taking into account the
commutation relation for the input field operators $\left[  \widehat{a}%
_{1,2}\left(  z,\omega\right)  ,\widehat{a}_{1,2}^{\dagger}\left(
z,-\omega^{\prime}\right)  \right]  =L/\left(  2\pi c\right)  \delta\left(
\omega+\omega^{\prime}\right)  $ in Eqs. (\ref{16b First Order CF}) and
(\ref{16c Two Photon CF}), we obtain%
\end{subequations}
\begin{subequations}
\begin{align}
G_{1,2}^{\left(  1\right)  }\left(  0\right)   &  =\frac{L}{2\pi c}\int
e^{-i\delta\tau}\left\vert B_{1,2}\right\vert ^{2}d\delta
,\label{17a solution 1st CF}\\
\Phi_{21}\left(  \tau\right)   &  =\frac{L}{2\pi c}\int e^{-i\delta\tau}%
A_{2}^{\ast}B_{1}d\delta. \label{17b}%
\end{align}

In particular, we are interested in the biphoton coherence time which is
determined by the width of the function $\Phi_{21}\left(  \tau\right)  $.
Harris and coworkers \cite{NBHarris} have pointed out that a long coherence
time can be obtained due to the effect of slow light \cite{HarrisPRA1992}
experienced by one of the photons of the entangled pair but not by the other.
Here we demonstrate that a long coherence time can be obtained even in a
optically thin medium%
\end{subequations}
\begin{equation}
\alpha_{1,2}\widetilde{L},~\beta_{1,2}\widetilde{L}\ll1,
\label{18 OptThinConditions}%
\end{equation}
where the time delay between the photons due to the slow light effect is negligible.

Under the conditions of Eq. (\ref{18 OptThinConditions}), Eqs.
(\ref{12a AB const}) and (\ref{12b}) simplify to%
\begin{equation}
A_{1,2}=1,~B_{1,2}=i\beta_{1,2}\widetilde{L}. \label{19 Simplyfied Constants}%
\end{equation}

To find the analytical form of the correlation function, we substitute Eqs.
(\ref{19 Simplyfied Constants}) into Eqs. (\ref{17a solution 1st CF}) and
(\ref{17b}) and integrate over $\delta$:%
\begin{subequations}
\begin{align}
G_{1,2}^{\left(  1\right)  }\left(  0\right)   &  =\frac{\zeta^{2}L^{3}}{2\pi
c}\int e^{-i\delta\tau}\left\vert \beta_{1,2}\right\vert ^{2}d\delta
\nonumber\\
&  \approx\frac{\zeta^{2}L^{3}\kappa^{2}\gamma_{ca}}{c\left(  1+\kappa\right)
^{3}\left(  \Gamma_{ba}^{2}+\Omega^{2}\right)  },\label{20a One Ph sol}\\
\Phi_{21}\left(  \tau\right)   &  =\frac{i\zeta L^{2}}{2\pi c}\int
e^{-i\delta\tau}\beta_{1}d\delta\nonumber\\
&  \approx\eta\int e^{-i\delta\tau}\left(  \frac{1}{W-i\delta}-\frac{1}%
{\Gamma_{ba}-i\left(  \Omega+\delta\right)  }\right)  d\delta\nonumber\\
&  =2\pi\eta\left(  e^{-W\left\vert \tau\right\vert }-e^{i\Omega\left\vert
\tau\right\vert }e^{-\Gamma_{ba}\left\vert \tau\right\vert }\right)  ,
\label{20b}%
\end{align}
where the constant $\eta\equiv\frac{\zeta L^{2}\kappa\gamma_{ca}}{2\pi
c\left(  1+\kappa\right)  \left(  \Gamma_{ba}-i\Omega\right)  }$ and also we
assume that $\Gamma_{ba}\gg W$. The normalized second-order correlation
function is then given by%

\end{subequations}
\begin{align}
&  g_{21}^{\left(  2\right)  }\left(  \tau\right)  =\frac{G_{21}^{\left(
2\right)  }\left(  \tau\right)  }{G_{1}^{\left(  1\right)  }\left(  0\right)
G_{2}^{\left(  1\right)  }\left(  0\right)  }\approx1+\nonumber\\
&  +e^{-W\left\vert \tau\right\vert }\frac{1+e^{-2\Gamma_{ba}\left\vert
\tau\right\vert }-2\cos\left(  \Omega\tau\right)  e^{-\Gamma_{ba}\left\vert
\tau\right\vert }}{\left\vert \beta_{\delta=0}\right\vert ^{2}\widetilde
{L}^{2}}. \label{21NormCF}%
\end{align}

When Eq. (\ref{18 OptThinConditions}) holds, the denominator of the second
term of Eq. (\ref{21NormCF}) is much less than unity and as a result the
visibility $V_{\text{vis}}=\left(  g_{\max}^{\left(  2\right)  }-g_{\min
}^{\left(  2\right)  }\right)  /\left(  g_{\max}^{\left(  2\right)  }+g_{\min
}^{\left(  2\right)  }\right)  \approx1$.

\begin{figure}
[ptb]
\begin{center}
\includegraphics[
trim=0.000000in 0.000000in -0.074520in 0.000000in,
height=2.1075in,
width=2.911in
]%
{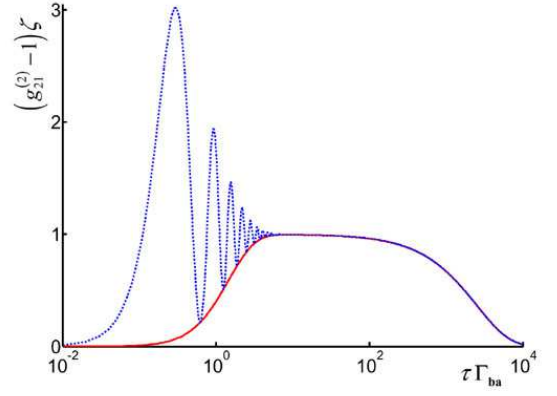}%
\caption{(color online) Normalized second-order correlation function for
$\kappa=1$ and $\gamma_{ca}/\Gamma_{ba}=10^{-4}$, for $\Omega=0$ (red solid
line) and $\Omega/\Gamma_{ba}=10$ (blue dotted line). The normalization
constant $\zeta\equiv\left\vert \beta_{\delta=0}\right\vert ^{2}\widetilde
{L}^{2}$ has the same value for both cases.}%
\label{fig2}%
\end{center}
\end{figure}

In Fig. (\ref{fig2}), we show the behavior of the normalized second-order
correlation function. It can be seen that the characteristic width is
determined by $1/W$ and also that there is an antibunching-like effect with a
characteristic time $1/\Gamma_{ba}$ (red solid line). When the pump detuning
is non-zero (blue dotted line), damped oscillations at frequency $\Omega$ with
a decay rate $1/\Gamma_{ba}$ are produced.%

This behavior is easy to understand if we look at Eq. (\ref{20b}), which shows
that the biphoton wave function is a coherent superposition of two different
FWM processes \cite{NBTLS}. The first term describes the generation of the
biphotons with a spectral width $W$ centered at the point $\delta=0$ (we call
this $FWM_{I}$) and the second term describes generation of the biphotons with
a spectral width $\Gamma_{ba}$ from the two sidebands which are centered at
the points $\delta=\pm\Omega$ (called $FWM_{II}$). As we can see from Eq.
(\ref{20b}), the phase shift between $FWM_{I}$ and $FWM_{II}$ oscillates at
the frequency of the pump detuning which leads to constructive or destructive
interference between them, as shown in Fig. (\ref{fig2}). The coherence time
of the biphoton generated by $FWM_{II}$ is determined by $1/\Gamma_{ba}$, so
that these biphotons can contribute to the coherent superposition only during
this period, which causes a fast damping of the oscillations. When the pump
detuning is zero, there is always destructive interference between the
biphotons from $FWM_{I}$ and $FWM_{II}$, leading to an anti-bunching dip with
a width $1/\Gamma_{ba}$ (we should note that the depth of the antibunching dip
never goes to zero as we have a macroscopic ensemble of the quantum systems).
The long coherence time is caused by biphotons that originate from the
$FWM_{I}$ process as it has a very narrow bandwidth [see Eq.
(\ref{08b CPO width})].

In summary, we have demonstrated that the combined effects of FWM and
long-lived CPOs in a TLS with intermediate metastable state are able to
produce narrowband biphotons with a long coherence time whose maximum value is
equal to the lifetime of the metastable state. The biphotons' waveform and
bandwidth can be controlled by the pump intensity. During the time
$1/\Gamma_{ba}$ the biphoton wavefunction shows antibunching behavior. If the
pump field is detuned, damped oscillation during this period is observed.

The authors thank A. Pe'er and A. Eilam for helpful discussions.

\end{document}